# Enhanced self-phase modulation in silicon nitride waveguides integrated with 2D graphene oxide films

Yuning Zhang, Jiayang Wu, *Member, IEEE,* Yunyi Yang, Yang Qu,
Houssein El Dirani, Romain Crochemore, Corrado Sciancalepore, Pierre Demongodin,
Christian Grillet, Christelle Monat, *Fellow, Optica (Formerly OSA)*,
and David J. Moss, *Fellow, IEEE, Fellow Optica (Formerly OSA)*

*Abstract*—**We experimentally demonstrate enhanced self-phase modulation (SPM) in silicon nitride (Si₃N₄) waveguides integrated with 2D graphene oxide (GO) films. GO films are integrated onto Si₃N₄ waveguides using a solution-based, transfer-free coating method that enables precise control of the film thickness. Detailed SPM measurements are carried out using both picosecond and femtosecond optical pulses. Owing to the high Kerr nonlinearity of GO, the hybrid waveguides show significantly improved spectral broadening compared to the uncoated waveguide, achieving a broadening factor of up to ~3.4 for a device with 2 layers of GO. By fitting the experimental results with theory, we obtain an improvement in the waveguide nonlinear parameter by a factor of up to 18.4 and a Kerr coefficient ($n_2$) of GO that is about 5 orders of magnitude higher than Si₃N₄. Finally, we provide a theoretical analysis for the influence of GO film length, coating position, and its saturable absorption on the SPM performance. These results verify the effectiveness of on-chip integrating 2D GO films to enhance the nonlinear optical performance of Si₃N₄ devices.**

Manuscript received XXX X, XXXX; revised XXX X, XXXX; accepted XXX X, XXXX. Date of publication XXX X, XXXX; date of current version XXX X, XXXX. This work was supported by the Australian Research Council Discovery Projects Programs (No. DP150102972 and DP190103186), the Swinburne ECR-SUPRA program, the Industrial Transformation Training Centers scheme (Grant No. IC180100005), and the Beijing Natural Science Foundation (No. Z180007). *(Corresponding author: Jiayang Wu, Baohua Jia, and David J. Moss.)*

Y. N. Zhang, J. Y. Wu, Y. Qu, and D. J. Moss are with Optical Sciences Center, Swinburne University of Technology, Hawthorn, VIC 3122, Australia. (e-mail:  yuningzhang@swin.edu.au, jiayangwu@swin.edu.au, yqu@swin.edu.au, dmoss@swin.edu.au).
H. El Dirani, R. Crochemore, and C. Sciancalepore are with CEA-LETI, Minatec, Optics ans Photonics Divesion, University Grenoble Alpes, 17 rue des Martyrs, 38054 Grenoble, France. (e-mail: houssein.eldirani@cea.fr, romain.crochemore@cea.fr, corrado.sciancalepore@soitec.com)
P. Demongodin, C. Grillet and C. Monat are with Institut des nanotechnologies de Lyon, UMR CNRS 5270, Ecole Centrale Lyon, F-69130 Ecully, France. (e-mail: pierre.demongodin@ec-lyon.fr, christian.grillet@ec-lyon.fr, christelle.monat@ec-lyon.fr)
Y. Y. Yang and B. H. Jia are with Center for Translational Atomaterials, Swinburne University of Technology, Hawthorn, VIC 3122, Australia. (e-mail: yunyiyang@swin.edu.au, bjia@swin.edu.au)
Color versions of one or more of the figures in this letter are available online at http://ieeexplore.ieee.org.
Digital Object Identifier 

## I. INTRODUCTION

Self-phase modulation (SPM) is a fundamental third-order nonlinear optical process that occurs when an optical pulse travelling in a nonlinear medium, where a varying refractive index of the medium is induced by the Kerr effect, thus produces a phase shift that leads to a change in the pulse's spectrum [1-3]. It has been widely used as a relevant all-optical modulation technology for a variety of applications in broadband optical sources [4, 5], optical spectroscopy [6, 7], pulse compression [8, 9], optical logic gates [10, 11], optical modulators / switches [12, 13], optical diodes [14, 15], and optical coherence tomography [16, 17].

The ability to realize SPM based on-chip integrated photonic devices will reap attractive benefits of compact footprint, high stability, high scalability, and low-cost mass production [18-21]. Although silicon (Si) has been a dominant device platform for integrated photonics [22-24], its strong two-photon absorption (TPA) at near-infrared wavelengths leads to a low nonlinear figure-of-merit (FOM = $n_2 / (\lambda \beta_{TPA})$, where $n_2$ is the Kerr nonlinearity, $\beta_{TPA}$ is the two photon absorption coefficient, and $\lambda$ the wavelength) of ~0.3 [25], which significantly limits the SPM performance of Si devices in the telecom band. To address this, other complementary metal-oxide-semiconductor (CMOS) compatible integrated platforms such as silicon nitride (Si₃N₄) and high-index doped silica glass (Hydex) have been exploited for nonlinear optics due to their negligible TPA in this wavelength range, which yields nonlinear FOMs >>1 [26-28]. Nevertheless, their low intrinsic Kerr nonlinearity ($n_2 = $ ~2.6 × 10⁻¹⁹ m² W⁻¹ and ~1.3 × 10⁻¹⁹ m² W⁻¹ for Si₃N₄ and Hydex, respectively, over an order of magnitude lower than Si [27, 29]) still poses a fundamental limitation with respect to the nonlinear efficiency [30, 31].

Recently, the on-chip integration of two-dimensional (2D) materials with ultrahigh Kerr nonlinearity has proven to be an effective way to overcome the limitations of these existing platforms and improve their nonlinear optical performance [32-35]. Enhanced SPM has been demonstrated in integrated waveguides incorporating graphene [36-38], MoS₂ [39], WS₂ [40], and graphene oxide (GO) [41]. Amongst the different 2D materials, GO has shown many advantages for implementing hybrid integrated photonic devices with superior SPM performance, including a large Kerr nonlinearity (about 4 orders of magnitude higher than Si [42, 43]), relatively low loss compared to other 2D materials (over 2 orders of magnitude lower than graphene [44, 45], facile synthesis processes [46, 47], and high compatibility with CMOS fabrication [48, 49]. In our previous work [41], we demonstrated enhanced SPM of picosecond optical pulses in Si waveguides integrated with 2D GO films, achieving a maximum spectral broadening factor (BF) of ~4.3 and enhanced FOM by up to 20 times.

In this paper, we demonstrate significantly improved SPM performance for Si₃N₄ waveguides integrated with 2D GO films. By using a solution-based, transfer-free coating method, we achieve on-chip integration of GO films with precise control of their thicknesses. We perform SPM measurements using both picosecond and femtosecond optical pulses centered at telecom wavelengths. Compared to the uncoated Si₃N₄ waveguide, the GO-coated waveguides show more significant spectral broadening for both the picosecond and femtosecond optical pulses, achieving a maximum BF of ~3.4 for a device with 2 layers of GO. We also fit the SPM experimental results with theory and obtain a Kerr coefficient ($n_2$) for GO that is about 5 orders of magnitude higher than Si₃N₄. Finally, we discuss the influence of GO film's length, coating position, and saturable absorption on the SPM performance. For Si waveguides the main challenge is to enhance the nonlinear FOM, whereas for Si₃N₄ waveguides the challenge is to enhance the nonlinear parameter $\gamma (= 2\pi n_2 / (\lambda A_{eff}))$, where $A_{eff}$ is the effective mode area) since their FOM is very large already. We obtain an enhancement in $\gamma$ by a factor of up to ~18.4 for a Si₃N₄ waveguide coated with 2 layers of GO, compared to the uncoated waveguide, accompanied by only a modest increase in the linear loss of about 3 dB/cm per layer of GO and with no measurable decrease in the nonlinear FOM. These results confirm the high nonlinear optical performance of Si₃N₄ waveguides integrated with 2D GO films.

## II. DEVICE FABRICATION AND CHARACTERIZATION

Fig. 1(a) shows a schematic of a GO-coated Si₃N₄ waveguide with a monolayer GO film. The bare Si₃N₄ waveguide has a cross-section of 1.6 μm × 0.66 μm, which was fabricated via a CMOS-compatible annealing-free and crack-free method [50, 51]. First, two-step deposition of Si₃N₄ film (330-nm-thick layer in each step) was achieved via low-pressure chemical vapor deposition (LPCVD) for strain management and crack prevention. Next, 248-nm deep ultraviolet lithography and CF₄/CH₂F₂/O₂ fluorine-based dry etching were employed for patterning the low-loss Si₃N₄ waveguides. A silica upper cladding was then deposited using high-density plasma-enhanced chemical vapor deposition (HDP-PECVD), followed by opening a window on it down to the top surface of the Si₃N₄ waveguide via lithography and reactive ion etching (RIE). Finally, the 2D layered GO film was coated onto the Si₃N₄ waveguide by using a solution-based method that enabled transfer-free and layer-by-layer film coating, as reported previously [45-47, 49]. Compared to the sophisticated film transfer processes employed for on-chip integration of other 2D materials such as graphene and TMDCs [38, 39, 52], our GO coating method is highly scalable, enabling precise control of the GO layer number (i.e., film thickness), large-area film coating, and good film attachment on integrated chips [43, 47]. Figs. 1(b-i) and (b-ii) show a schematic cross section and the transverse electric (TE) mode profile of the GO-coated Si₃N₄ waveguide in Fig. 1(a), respectively. The interaction between light and the GO film possessing an ultrahigh Kerr nonlinearity can be excited by the waveguide evanescent field, which underpins the enhancement of the SPM response in the hybrid waveguide.

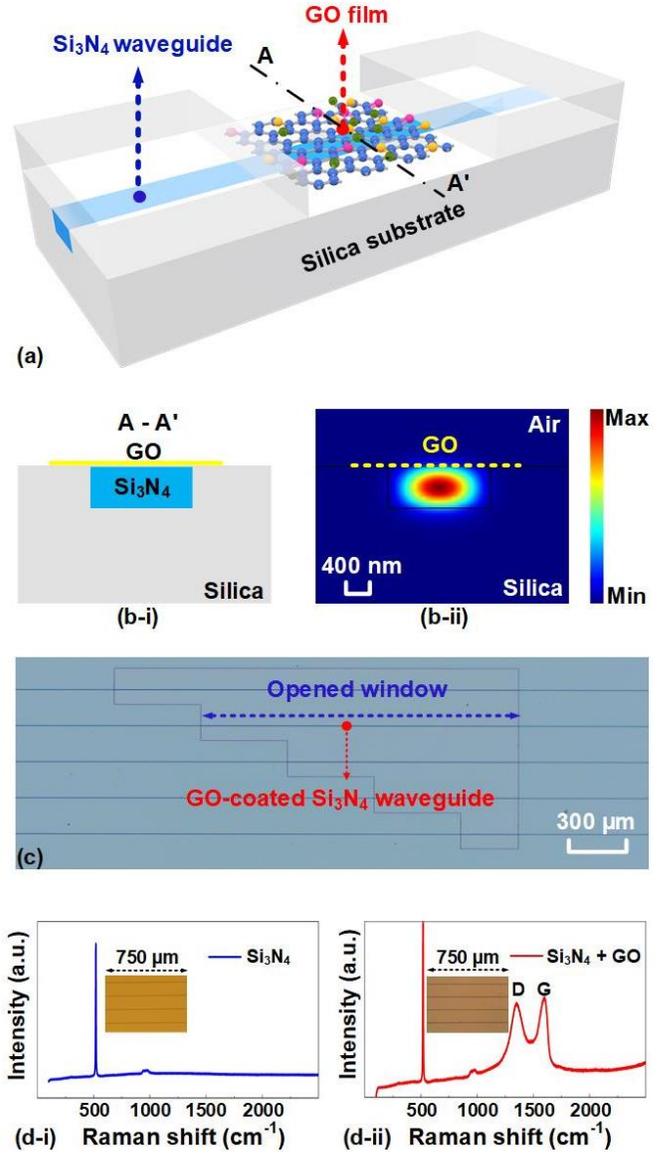

Fig. 1. (a) Schematic illustration of a GO-coated Si₃N₄ waveguide with a monolayer GO film. (b-i) Schematic illustration of cross section and (b-ii) corresponding TE mode profile of the GO-coated Si₃N₄ waveguide in (a). (c) Microscope image of a Si₃N₄ integrated chip uniformly coated with monolayer GO film. (d) Raman spectra of a Si₃N₄ chip (i) before and (ii) after coating 2 layers of GO. Insets show the corresponding microscope images.

Fig. 1(c) shows a microscope image of a Si₃N₄ integrated chip uniformly coated with a monolayer GO film, where the coated GO film exhibits good morphology, high transmittance, and high uniformity. The opened window on the silica upper cladding of the uncoated Si₃N₄ chip enables control of the film length and placement of the GO film that are in contact with the $Si_3N_4$ waveguide. Note that this can also be realized by patterning GO films on planarized $Si_3N_4$ waveguides (without silica upper cladding) via lithography and lift-off processes, as we did in our previous work [44]. In this work, we used $Si_3N_4$ waveguides with opened windows mainly because they have lower coupling loss and propagation loss, which is beneficial for boosting the nonlinear response of SPM.

Figs. 1(d-i) and (d-ii) show the measured Raman spectra of a $Si_3N_4$ chip before and after coating 2 layers of GO, respectively, where the presence of the representative D and G peaks of GO in the latter one verifies the successful on-chip integration of the GO film [44, 46]. According to our previous measurements [41, 44, 45], the GO film thickness shows a near linear relationship with layer number at small film thickness (i.e., layer numbers < 100), and the thickness for 1 layer of GO is ~2.0 nm. For the GO-coated $Si_3N_4$ waveguides used in the following SPM measurements, the measured film thicknesses for 1 and 2 layers of GO are ~2.1 nm and ~ 4.3 nm, respectively.

## III. LOSS MEASUREMENTS

Fig. 2 shows the experimental setup used for measuring both loss and SPM of GO-coated $Si_3N_4$ waveguides. Three different laser sources were employed, including a tunable continuous-wave (CW) laser and two different fiber pulsed lasers (FPLs) that can generate nearly Fourier-transform limited picosecond (pulse duration: ~1.9 ps) and femtosecond optical pulses (pulse duration: ~180 fs) centered at telecom wavelengths. An optical isolator was inserted after the laser source to prevent the reflected light from damaging it. A variable optical attenuator (VOA) and a polarization controller (PC) were used to tune the power and polarization of the input light, respectively. For both the loss and SPM measurements, TE polarization of input light injected into the device under test (DUT) was chosen because it supports in-plane interaction between the waveguide evanescent field and the 2D GO film, which is much stronger compared to the out-of-plane interaction given the significant optical anisotropy of 2D materials [32, 45]. We used inverse-taper couplers at both ends of the $Si_3N_4$ waveguide, which were butt coupled to lensed fibers to achieve light coupling into and out of the DUT.

For the loss measurements, the power of the light before and after passing the DUT was measured by two optical power meters, i.e., OPM 1 and OPM 2. All the three laser sources were used to measure the loss of bare and GO-coated $Si_3N_4$ waveguides. The corresponding results are compared in Fig. 3.

Fig. 3(a) shows the insertion loss ($IL_{CW}$) of GO-coated $Si_3N_4$ waveguides versus input CW light power. Unless otherwise specified, the input power of CW light or optical pulses in this paper represents the power coupled into the waveguide after excluding the fiber-to-chip coupling loss. We measured the hybrid waveguides with 1 and 2 layers of GO (i.e., layer number $N$ = 1, 2), and the corresponding results for the uncoated $Si_3N_4$ waveguide ($N$ = 0) are also shown for comparison. The total length of the $Si_3N_4$ waveguide was 20 mm and the length of the opened window was 1.4 mm. The opened window started at 0.7 mm after the light input port.

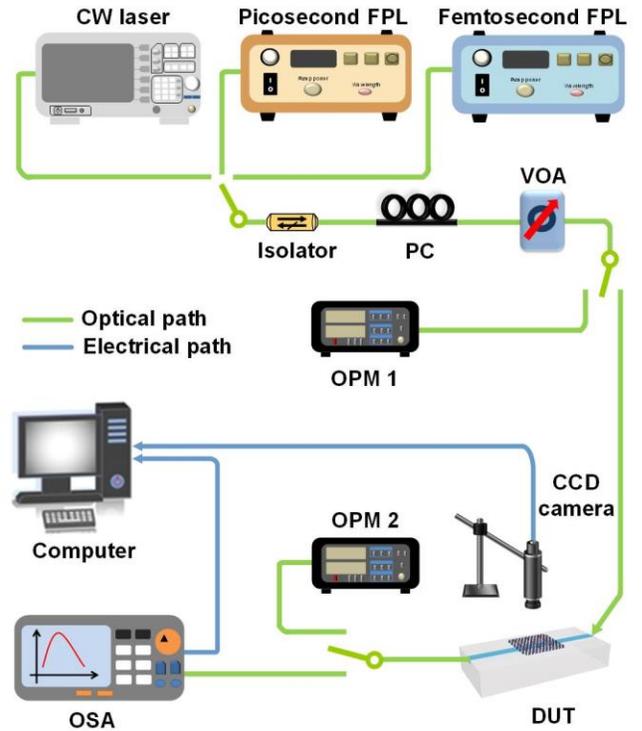

Fig. 2. Experimental setup for measuring loss and SPM of GO-coated $Si_3N_4$ waveguides. CW laser: continuous-wave laser. FPL: fiber pulsed laser. PC: polarization controller. VOA: variable optical attenuator. OPM: optical power meter. DUT: device under test. CCD: charged-coupled device. OSA: optical spectrum analyzer.

As can be seen, the insertion loss does not show any obvious variation with the power of the input CW light, reflecting that the power-dependent loss induced by the photo-thermal changes in the GO films is negligible. This is because the photo-thermal changes are sensitive to the average light power coupling into the GO-coated waveguides [44, 48], and the average power of the input CW light here (< 7 mW) is much lower than those inducing significant photo-thermal changes (> 40 mW) in previous work [44, 48, 53].

According to the results in Fig. 3(a), the excess propagation losses induced by the GO films are ~3.0 dB/cm and ~6.1 dB/cm for the hybrid waveguides with 1 and 2 layers of GO, respectively. These values are slightly higher than those of GO-coated dope silica waveguides [45, 46] but lower than those of GO-coated Si waveguides [41], mainly due to the moderate GO mode overlap in the GO-coated $Si_3N_4$ waveguides. It is also worth mentioning that the GO-induced excess propagation loss is about 100 times lower than graphene-induced excess propagation loss in graphene-coated $Si_3N_4$ waveguides [51, 54], highlighting the low material absorption of GO compared to graphene and its advantage for implementing nonlinear photonic devices with relatively low loss. The low loss of GO is mainly induced by its large bandgap, which is typically between 2.1 eV – 3.6 eV [43, 55, 56]. In principle, GO with a bandgap > 2 eV has negligible linear absorption below its bandgap, e.g., at near-infrared wavelengths (with a photon energy of ~0.8 eV at 1550 nm). The light absorption of practical GO films is mainly caused by defects as well as scattering loss stemming from imperfect layer contact and film unevenness [41, 44, 45].

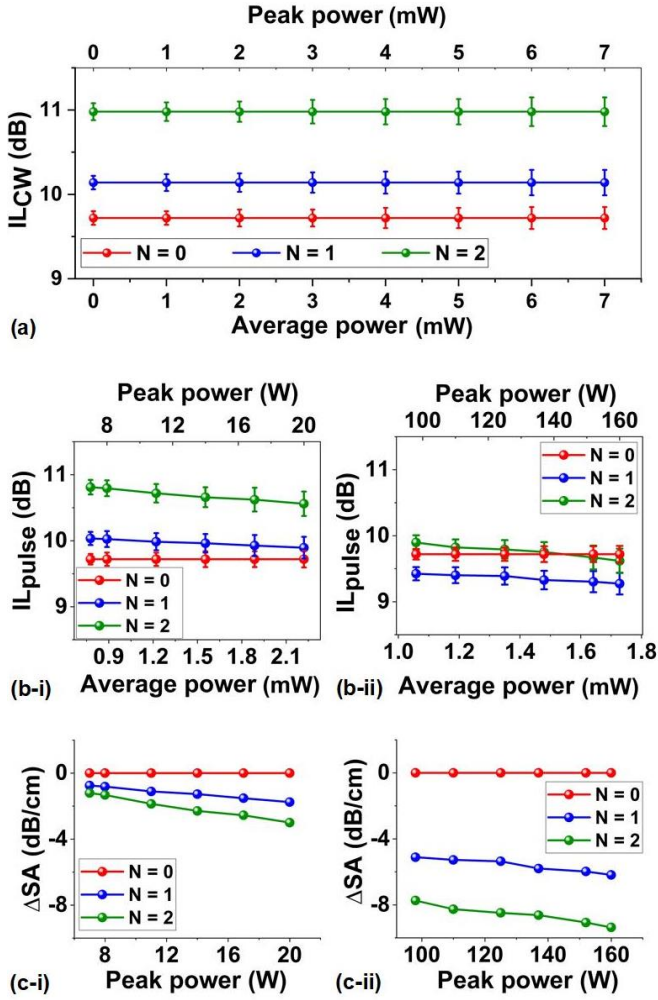

Fig. 3. (a) Measured insertion loss ($IL_{CW}$) of GO-coated Si₃N₄ waveguides versus input power of continuous-wave (CW) light. (b) Measured insertion loss ($IL_{pulse}$) of GO-coated Si₃N₄ waveguides versus input power of optical pulses. (c) Excess propagation loss induced by the SA ($\Delta SA$) versus peak power of input optical pulses. In (b) and (c), (i) and (ii) show the results for picosecond and femtosecond optical pulses, respectively. In (a) − (c), the results for uncoated ($N = 0$) and hybrid Si₃N₄ waveguides coated with 1 and 2 layers of GO ($N = 1, 2$) are shown for comparison. The data points depict the average of measurements on three samples and the error bars illustrate the variations among the different samples.

Figs. 3(b-i) and 3(b-ii) show the insertion loss ($IL_{pulse}$) of GO-coated Si₃N₄ waveguides versus input power of picosecond and femtosecond optical pulses, respectively. In contrast to CW light that has a peak power equaling to its average power, picosecond and femtosecond optical pulses have peak powers that are much higher than their average powers. Both the picosecond and femtosecond FPLs we used had the same repetition rate of ~60 MHz. For the picosecond pulses, the average input power ranged between 0.8 mW and 2.3 mW, which corresponded to a peak power range of 7 W – 20 W. For the femtosecond pulses, the average input power ranged between 1.1 mW and 1.7 mW, which corresponded to a peak power range of 98 W – 160 W. Given that the average powers of the picosecond and femtosecond optical pulses are in the same level as that of the CW light in Fig. 3(a), the photo-thermal changes in the GO films can be neglected when these optical pulses go through the hybrid waveguides.

In both Figs. 3(b-i) and 3(b-ii), the measured $IL_{pulse}$ of GO-coated Si₃N₄ waveguides decreases with the input power of optical pulses, and the waveguide with 2 layers of GO shows a more obvious decrease than the waveguide with 1 layer of GO. In contrast, the result for the uncoated Si₃N₄ waveguide does not show such a trend. This indicates that there is saturable absorption (SA) induced by the GO films in the hybrid waveguides. Similar phenomenon has also been observed for the GO-coated Si waveguides [41] and graphene-coated Si₃N₄ waveguides [51]. In our experiment, we also note that the change in the loss of the hybrid waveguides was not permanent, and the measured $IL_{pulse}$ in Fig. 3(b) is repeatable.

Fig. 3(c) depicts the SA-induced excess propagation loss ($\Delta SA$, after excluding the linear propagation loss) versus the peak power of the input optical pulses, which is extracted from the results in Figs. 3 (a) and (b). The negative values of $\Delta SA$ indicate that the loss decreases with light power, showing an opposite trend to TPA where the loss increases with light power [25, 33, 57]. The decreased loss induced by SA can facilitate more significant SPM driven by a high optical power. For femtosecond pulses, the decrease in loss is more significant than that for picosecond pulses, which can be attributed to their relatively high peak power that induces more significant SA in the GO.

## IV. SPM MEASUREMENTS

In the SPM measurements, we used the same FPLs and the same fabricated devices as those employed for loss measurements in Section III to measure the SPM-induced spectral broadening. As shown in Fig. 2, picosecond or femtosecond optical pulses generated by these FPLs were coupled into the DUT, and the output signal was sent to an optical spectrum analyzer (OSA) for observation of spectral broadening. The corresponding results for the picosecond and femtosecond optical pulses are compared in Figs. 4 and 5, respectively.

Fig. 4(a) shows the normalized spectra of picosecond optical pulses before and after propagation through the uncoated and GO-coated Si₃N₄ waveguides. The peak power of the input picosecond optical pulses was kept the same at ~20 W. The output spectrum from the uncoated Si₃N₄ waveguide shows slight spectral broadening compared to the input pulse spectrum, which is mainly induced by the SPM in the Si₃N₄ waveguide. In contrast, the output spectra after propagating through the GO-coated Si₃N₄ waveguides show more significant spectral broadening, reflecting the enhanced SPM in these hybrid waveguides.

Fig. 4(b) shows the output spectra after propagation through the hybrid waveguide with 2 layers of GO measured using picosecond optical pulses with different peak powers. We chose 6 different peak powers ranging from 7 W to 20 W – the same as those in Fig. 3(b-i). As expected, the spectral broadening of the output spectra becomes more significant as the peak power increases.

To quantitively compare the spectral broadening in these waveguides, we calculated the BFs for the measured output spectra. The BF is defined as [37, 41, 51]:

$$BF = \frac{\Delta \omega_{rms}}{\Delta \omega_0} \qquad (1)$$

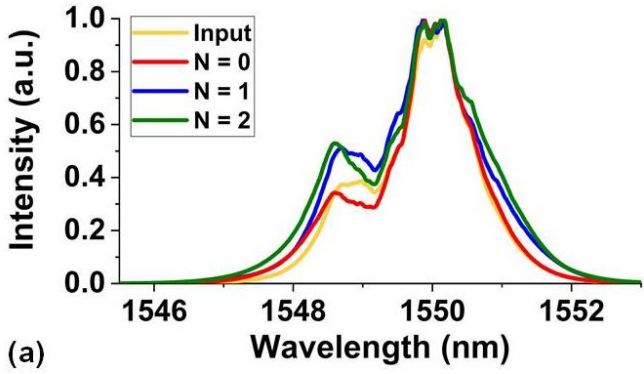

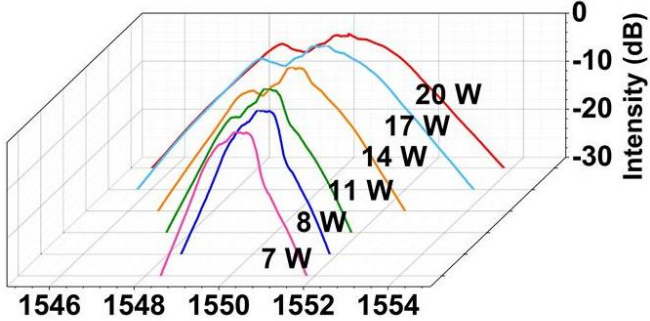

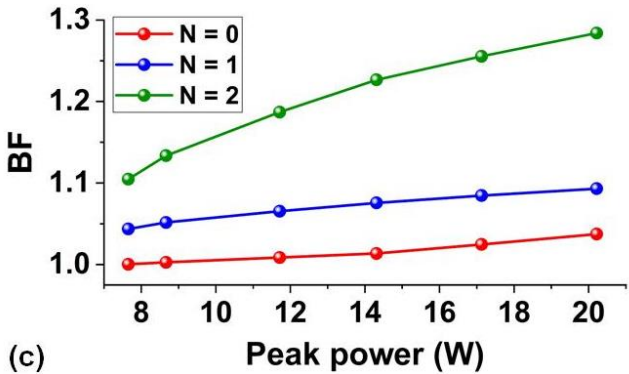

Fig. 4. SPM experimental results using picosecond optical pulses. (a) Normalized spectra of optical pulses before and after propagation through the GO-coated $Si_3N_4$ waveguides with 1 and 2 layers of GO at an input peak power of ~20 W. (b) Optical spectra measured at different input peak powers for the hybrid waveguides with 2 layers of GO. (c) BFs of the measured output spectra versus input peak power for the hybrid waveguides with 1 and 2 layers of GO. In (a) and (c), the corresponding results for the uncoated $Si_3N_4$ waveguides are also shown for comparison.

where $\Delta\omega_0$ and $\Delta\omega_{rms}$ are the root-mean-square (RMS) spectral widths of the input and output signals, respectively.

Fig. 4(c) shows the BFs for the uncoated and GO-coated $Si_3N_4$ waveguides versus the peak power of input picosecond optical pulses. As can be seen, the BFs for the GO-coated $Si_3N_4$ waveguides are higher than that of the uncoated waveguide, and the BF for the hybrid waveguide with 2 layers of GO is higher than that for the device with 1 layer of GO, showing agreement with the results in Fig. 4(a). The BF increases with the peak power of the optical pulses, which is consistent with the results in Fig. 4(b). At a peak power of ~20 W, a maximum BF of ~1.3 is achieved for the hybrid waveguide with 2 layers of GO.

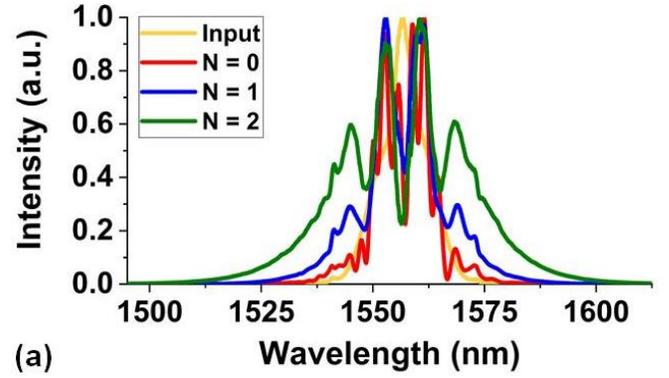

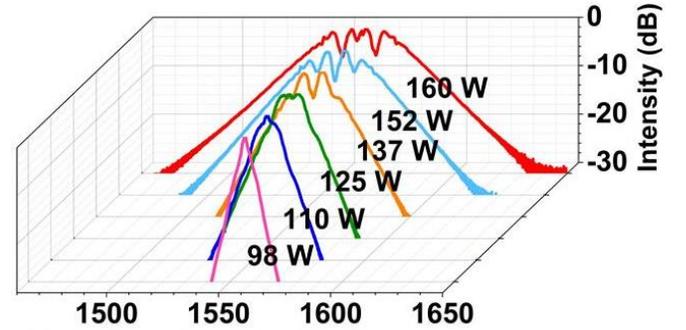

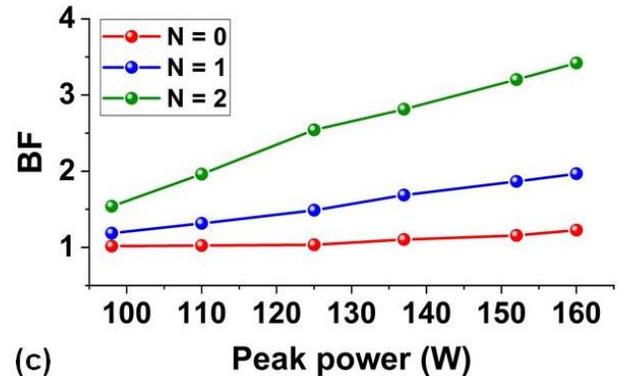

Fig. 5. SPM experimental results using femtosecond optical pulses. (a) Normalized spectra of optical pulses before and after propagation through the GO-coated $Si_3N_4$ waveguides with 1 and 2 layers of GO at an input peak power of ~160 W. (b) Optical spectra measured at different input peak powers for the hybrid waveguides with 2 layers of GO. (c) BFs of the measured output spectra versus input peak power for the hybrid waveguides with 1 and 2 layers of GO. In (a) and (c), the corresponding results for the uncoated $Si_3N_4$ waveguides are also shown for comparison

Fig. 5(a) shows the normalized spectra of femtosecond optical pulses before and after propagation through the uncoated and GO-coated $Si_3N_4$ waveguides, which were measured at the same input peak power of ~160 W. Similar to Fig. 4(a), the output spectra after passing through the hybrid waveguides show more significant spectral broadening compared to the uncoated waveguide. Fig. 5(b) shows the output spectra measured at different input peak powers for the hybrid waveguide with 2 layers of GO, showing the similar trend as that in Fig. 4(b). The peak power of the input femtosecond optical pulses ranged from 98 W to 160 W – the same as those in Fig. 3(b-ii). The calculated BFs for the uncoated and GO-coated $Si_3N_4$ waveguides versus the input peak power are shown in Fig. 5(c). A maximum BF of ~3.4 is achieved at a peak power of 160 W for the hybrid waveguide with 2 layers of GO, which is ~2.6 times higher than the maximum BF achieved for the picosecond optical pulses. This mainly results from the relatively high peak power of the femtosecond optical pulses that drives more significant SPM in the hybrid waveguide.

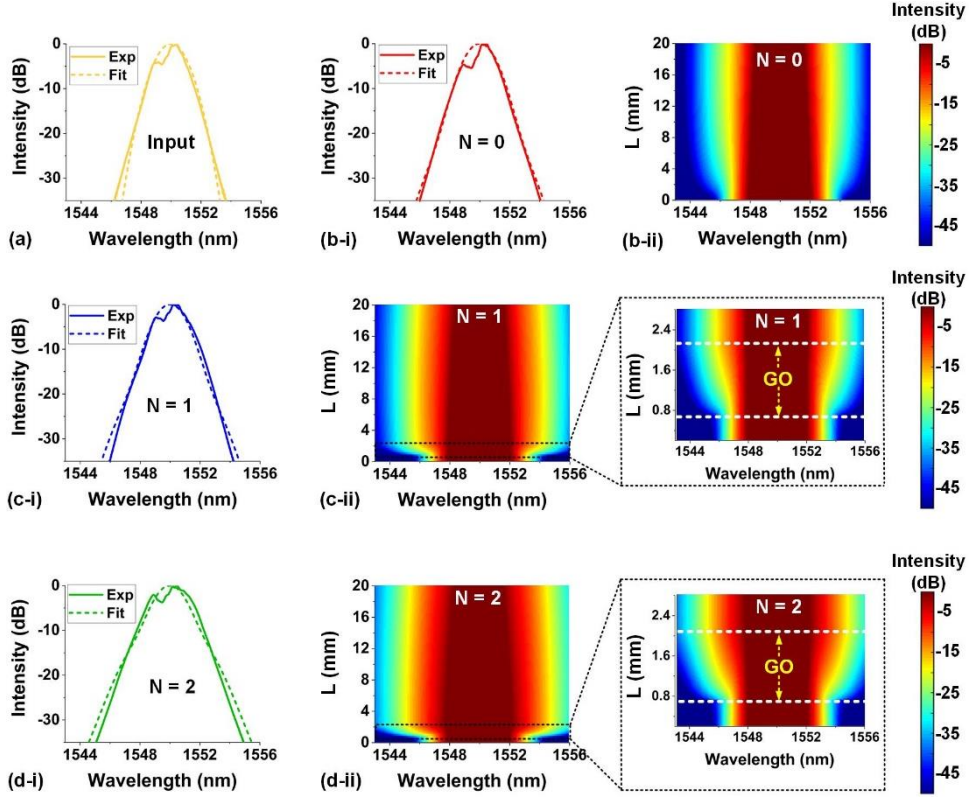

Fig. 6. (a) Measured and fit spectra of input picosecond optical pulses. (b-i) Measured and fit output spectra after propagation through the uncoated $Si_3N_4$ waveguides. (b-ii) Simulated spectra evolution along the uncoated $Si_3N_4$ waveguide. (c-i) Measured and fit output spectra after propagation through the hybrid waveguide with 1 layer of GO. (c-ii) Simulated spectra evolution along the hybrid waveguide with 1 layer of GO. (d-i) Measured and fit output spectra after propagation through the hybrid waveguide with 2 layers of GO. (d-ii) Simulated spectra evolution along the hybrid waveguide with 2 layers of GO. Insets in (c-ii) and (d-ii) show zoom-in views for the GO-coated regions. In (b) – (d), the peak power of the input picosecond pulses is ~20 W.

## V. THEORETICAL ANALYSIS AND DISCUSSION

Based on the theory in Refs. [21, 41, 58], we simulated the evolution of optical pulses traveling along the GO-coated $Si_3N_4$ waveguides using the nonlinear Schrodinger equation as follows:

$$\frac{\partial A}{\partial z} = -\frac{i\beta_2}{2}\frac{\partial^2 A}{\partial t^2} + i\gamma\,|A|^2A - \frac{1}{2}\alpha A \qquad (2)$$

where $i = \sqrt{1}$, $A(z, t)$ is the slowly varying temporal pulse envelope along the propagation direction $z$, $\beta_2$ is the second-order dispersion coefficient, and $\gamma$ is the waveguide nonlinear parameter. The overall loss factor $\alpha$ includes both the linear propagation loss and the SA-induced excess propagation loss that are discussed in Fig. 3.

Unlike in Refs. [41, 58], there are no free carrier absorption (FCA) and free carrier dispersion (FCD) items in Eq. (2) since the TPA in both $Si_3N_4$ and GO (with bandgaps > 2 eV) is negligible at near-infrared wavelengths. We retain only the second-order dispersion item in Eq. (2) because the physical length of the waveguides (20 mm) is much smaller than the dispersion length (> 1 m) [59]. In our simulation, we divided the GO-coated $Si_3N_4$ waveguides into uncoated (with silica cladding) and hybrid segments (coated with 1.4-mm-long GO films). Numerically solving Eq. (2) was performed for each segment, and the output from the previous segment was set as the input for the subsequent one.

Figs. 6(a) and (b-i) show the measured and fit spectra of the input picosecond pulses and the output signal after propagation through the uncoated $Si_3N_4$ waveguide, respectively. The simulated spectrum evolution of the input optical pulses propagating along the uncoated waveguide is shown in Fig. 6(b-ii). The peak power of the input picosecond pulses is ~20 W. The fit spectra and spectrum evolution were calculated based on Eq. (2), which show good agreement with the experimental results. The slight discrepancies between the measured and fit spectra mainly result from imperfections of the input pulse spectrum. The fit $\gamma$ for the uncoated $Si_3N_4$ waveguide is ~1.5 $W^{-1}m^{-1}$ – in agreement with the reported values in previous literature [20, 44, 60].

Figs. 6(c-i) and (d-i) show the measured and fit spectra for the output signals after transmission through the hybrid waveguides with 1 and 2 layers of GO, respectively. The input peak power is the same as that in Fig. 6(b). The corresponding spectrum evolutions along the hybrid waveguides are shown in Figs. 6(c-ii) and (d-ii). As can be seen, the theoretical

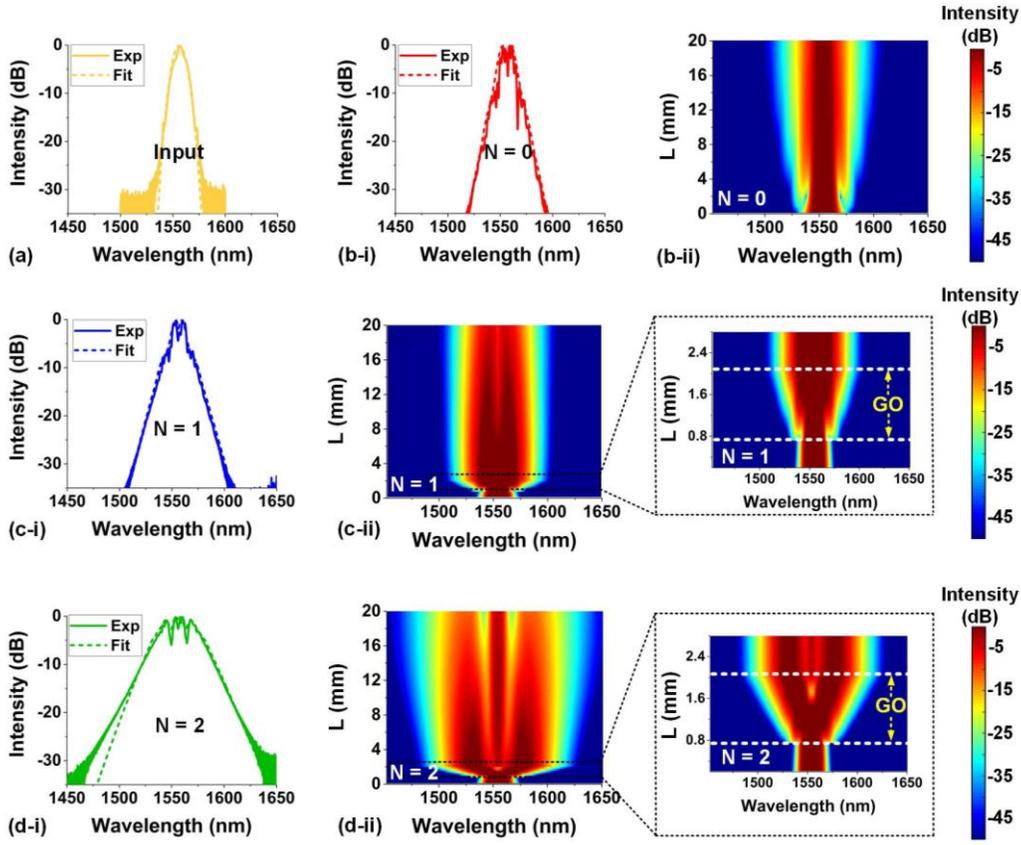

Fig. 7. (a) Measured and fit spectra of input femtosecond optical pulses. (b-i) Measured and fit output spectra after propagation through the uncoated Si₃N₄ waveguides. (b-ii) Simulated spectra evolution along the uncoated Si₃N₄ waveguide. (c-i) Measured and fit output spectra after propagation through the hybrid waveguide with 1 layer of GO. (c-ii) Simulated spectra evolution along the hybrid waveguide with 1 layer of GO. (d-i) Measured and fit output spectra after propagation through the hybrid waveguide with 2 layers of GO. (d-ii) Simulated spectra evolution along the hybrid waveguide with 2 layers of GO. Insets in (c-ii) and (d-ii) show zoom-in views for the GO-coated regions. In (b) – (d), the peak power of the input femtosecond pulses is ~160 W.

simulations also agree well with the experimental results. The fit $\gamma$'s for the hybrid waveguides with 1 and 2 layers of GO are ~11.5 and ~27.6, respectively, which are ~7.7 and ~18.4 times that of the uncoated Si₃N₄ waveguide, reflecting the significantly improved Kerr nonlinearity for the hybrid waveguides. The significant Kerr nonlinearity is also confirmed by the dramatical spectral broadening within the GO-coated region, as shown in the insets of Figs. 6(c-ii) and (d-ii).

Similar to Fig. 6, we also performed simulations based on Eq. (2) to fit the experimental results of femtosecond optical pulses. The measured and fit spectra of the input femtosecond pulses and the output signal after propagation through the uncoated Si₃N₄ waveguide are shown in Figs. 7(a) and (b-i) respectively. Figs. 7(c-i) and (d-i) show the measured and fit spectra for the output femtosecond signals after transmission through the hybrid waveguides with 1 and 2 layers of GO, respectively. The simulated spectrum evolution of the input pulses corresponding to Figs. 7(b-i), 7(c-i), and 7(d-i) are shown in Figs. 7(b-ii), 7(c-ii), and 7(d-ii), respectively. In all of these figures, the peak power of the input femtosecond pulses is ~160 W. As can be seen, all the theoretical curves show good agreement with the experimental ones. The fit $\gamma$ for the uncoated Si₃N₄ waveguide and the hybrid waveguides with 1 and 2 layers of GO are the same as those obtained by fitting the experimental results of picosecond optical pulses in Fig. 6, highlighting the high consistency and further confirming the enhanced Kerr nonlinearity for the hybrid waveguides. Similar to the insets of Figs. 6(c-ii) and (d-ii), there is also dramatical spectral broadening within the GO-coated region in the insets of Figs. 7(c-ii) and (d-ii).

Based on the fit $\gamma$'s of the hybrid waveguides, we further extract the Kerr coefficient ($n_2$) of the layered GO films using [46, 61, 62] :

$$\gamma = \frac{2\pi}{\lambda_c} \frac{\iint_D n_0^2(x,y) n_2(x,y) S_z^2 dxdy}{\left[ \iint_D n_0(x,y) S_z dxdy \right]^2} \tag{3}$$

where $\lambda_c$ is the pulse central wavelength, $D$ is the integral of the optical fields over the material regions, $S_z$ is the time-averaged Poynting vector calculated using mode solving software, $n_0(x, y)$ is the refractive index profiles calculated over the waveguide cross section and $n_2(x, y)$ is the Kerr coefficient of the different material regions. The values of $n_2$ for silica and Si₃N₄ used in our calculation were $2.60 \times 10^{-20}$ m² W⁻¹ [27] and $2.59 \times 10^{-19}$ m² W⁻¹, respectively, with the latter obtained by fitting the experimental results for the uncoated Si₃N₄ waveguide.

The fit $\gamma$'s of the hybrid waveguides with 1 and 2 layers of GO are ~11.5 W⁻¹m⁻¹and ~27.6 W⁻¹m⁻¹, respectively, which are ~7.7 and ~18.4 times that of the uncoated Si₃N₄ waveguide. The extracted $n_2$ of 1 and 2 layers of GO are ~1.23 × $10^{-14}$ m² $W^{-1}$ and ~1.19 × $10^{-14}$ m² $W^{-1}$, respectively. Both of the values are about 5 orders of magnitude higher than that of $Si_3N_4$ and agree reasonably well with our previous measurements [44]. Note that the $n_2$ of 1 layer of GO is higher than that of 2 layers of GO. We infer this may result from the increased inhomogeneous defects within the GO layers and imperfect contact between the multiple GO layers. Nonetheless, the higher GO mode overlap for the thicker 2-layer film, compared to the single-layer film, resulted in a more than doubling of the nonlinear parameter $\gamma$.

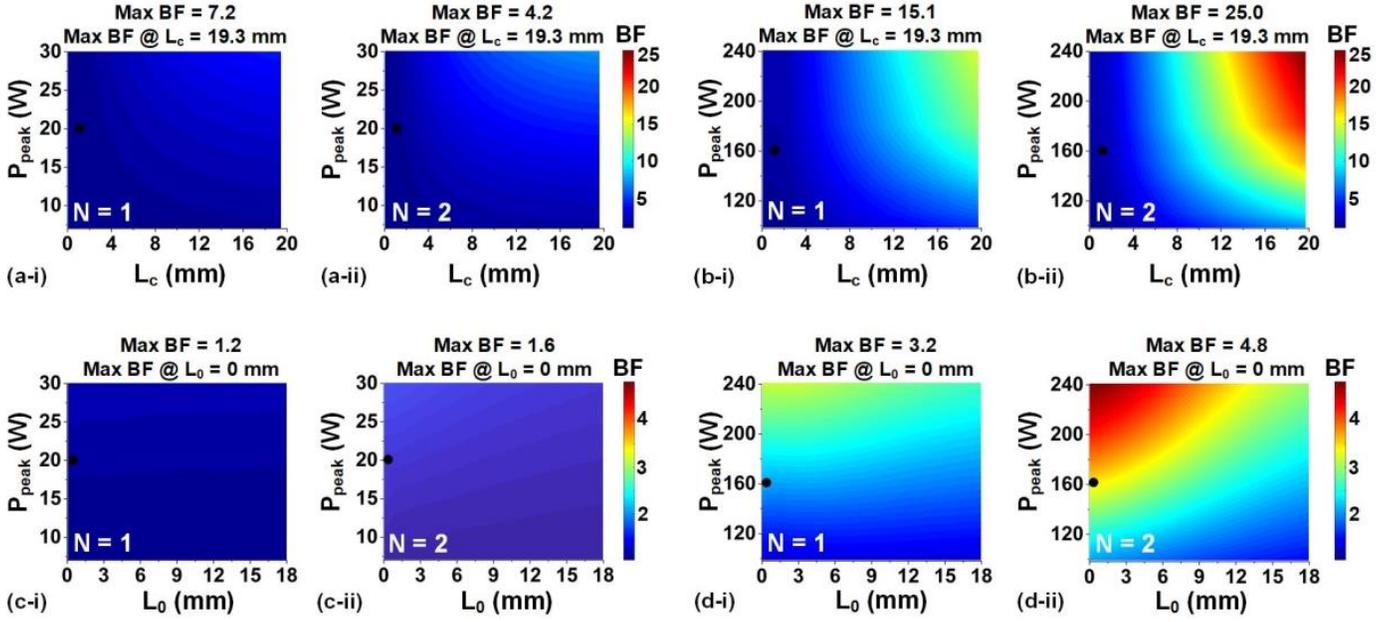

Fig. 8. (a) BFs versus GO film length ($L_c$) and input peak power ($P_{peak}$) for picosecond optical pulses after propagation through the hybrid waveguides. (b) BFs versus $L_c$ and $P_{peak}$ for femtosecond optical pulses after propagation through the hybrid waveguides. (c) BFs versus GO coating position ($L_0$) and $P_{peak}$ for picosecond optical pulses after propagation through the hybrid waveguides. (d) BFs versus $L_0$ and $P_{peak}$ for femtosecond optical pulses after propagation through the hybrid waveguides. In (a) – (d), (i) and (ii) show the corresponding results for the waveguides with 1 and 2 layers of GO, the black points mark the results corresponding to the device parameters and input powers in Figs. 4(a) and 5(a). In (a) and (b), $L_0 = 0.7$ mm. In (c) and (d), $L_c = 1.4$ mm.

Based on the SPM modeling in Eq. (2) and the fit parameters obtained from Figs. 6 and 7, we further investigate the influence of GO film length ($L_c$) and coating position ($L_0$) on the SPM performance of GO-coated $Si_3N_4$ waveguides.

Figs. 8(a) and (b) show the calculated BFs versus $L_c$ and input peak power ($P_{peak}$) for picosecond and femtosecond optical pulses after propagation through the hybrid waveguides, respectively. In each figure, (i) and (ii) show the results for the waveguides with 1 and 2 layers of GO, respectively. The coating position is fixed at $L_0 = 0.7$ mm – the same as those of the fabricated devices in Sections III and IV. The black points mark the parameters corresponding to the SPM measurements in Section IV, where the calculated BFs are consistent with the experimental results in Figs. 4 and 5. The BF increases with both $L_c$ and $P_{peak}$, with maximum BFs of 4.2 (at $L_c = 19.3$ mm and $P_{peak} = 30$ W) and 25.0 (at $L_c = 19.3$ mm and $P_{peak} = 240$ W) being achieved for the picosecond and femtosecond optical pulses, respectively. This reflects that there is a large room for improvement in the SPM-induced spectral broadening by increasing the GO film length and the input peak power. The BF can also be improved by coating thicker GO films (i.e., $N > 2$), which was used for increasing the FWM conversion efficiency in Ref. [44]. The increased GO film thickness will also lead to loss increase for the hybrid waveguides and hence creates a need to balance the trade-off between the Kerr nonlinearity and loss [30, 31].

Figs. 8(c) and (d) show the calculated BFs versus $L_0$ and $P_{peak}$ for picosecond and femtosecond optical pulses after propagation through the hybrid waveguides, respectively, where the film length is fixed at $L_c = 1.4$ mm. The simulation results marked by the black points also agree well with the experimental results in Figs. 4 and 5. The BF increases with $P_{peak}$ – a trend similar to that in Figs. 8(a) and (b). In contrast, it decreases with $L_0$, with the maximum value being achieved at $L_0 = 0$. This indicates that the largest spectral broadening can be achieved by coating GO films at the beginning, as expected since the light power is highest at the start of the waveguide.

As discussed in Section III, the decreased loss induced by the SA in the GO films affects the SPM performance. In Figs. 9(a) and (b), we show the influence of the SA on the spectral broadening of picosecond and femtosecond optical pulses after propagation through the hybrid waveguide with 2 layers of GO, respectively. In each figure, the solid curve shows the result when considering the SA that induces a slightly reduced loss, whereas the dashed curve shows the result that was calculated using a constant linear loss of GO measured at low CW powers (i.e., the loss in Fig. 3(a)). As can be seen, the SA of GO has a positive influence and yields more significant spectral broadening for both the picosecond and femtosecond optical pulses. The difference for the femtosecond pulses is more obvious, showing a similar trend to the results for the loss decrease in Fig. 3(b) and reflecting that there is a more significant influence of the SA on the spectral broadening for optical pulses with higher peak powers.

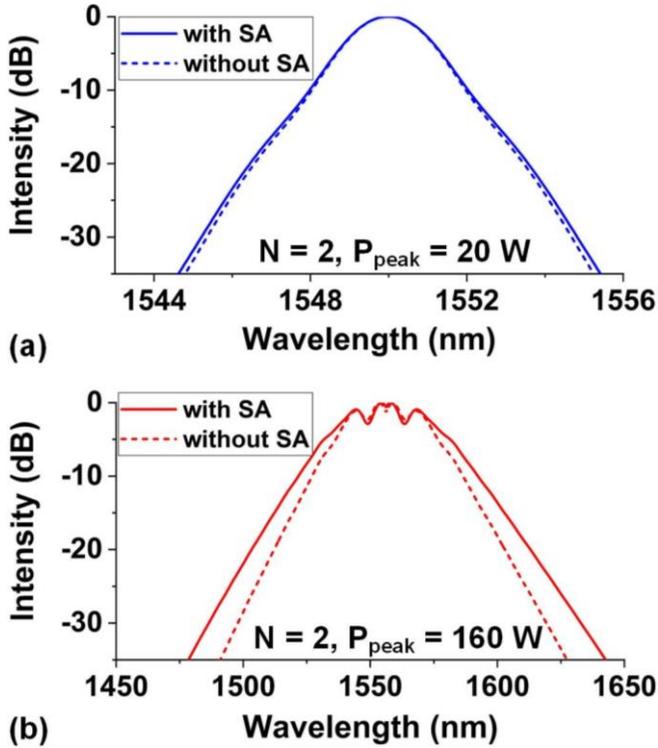

Fig. 9. (a) Comparison of spectral broadening of picosecond optical pulses after propagation through the GO-Si$_3$N$_4$ waveguide with and without considering the SA of GO. (b) Comparison of spectral broadening of femtosecond optical pulses after propagation through the GO-Si$_3$N$_4$ waveguide with and without considering the SA of GO. In (a) and (b), $N = 2$, $L_c = 1.4$ mm, and $L_0 = 0.7$ mm. The peak power for the picosecond and femtosecond optical pulses are 20 W and 160 W, respectively.

In Table I, we provide comparisons for the parameters of the GO-Si$_3$N$_4$ waveguides in this work and the GO-Si waveguides in Ref. [41]. We note that the trade-offs and challenges involved with integrating GO films into these two very different platforms, are in turn very different. Compared to the GO-Si waveguides, GO-Si$_3$N$_4$ waveguides have a larger waveguide geometry, which results in lower mode overlap with GO films. Such a reduced GO mode overlap yields a lower GO-induced excess propagation loss, at the expense of a weaker light-GO interaction. Despite this, the nonlinear parameter $\gamma$ of the GO-Si$_3$N$_4$ waveguide with 1 layer of GO is still ~7.7 times that of the uncoated waveguide. In contrast, there is only about a 2-fold improvement in the $\gamma$ of the GO-Si waveguide with 1 layer of GO. This mainly due to the relatively low $n_2$ of Si$_3$N$_4$ compared to Si, reflecting that integrating GO onto Si$_3$N$_4$ waveguides has a more dramatic impact on improving the nonlinear performance. In contrast to Si that has strong TPA at near infrared wavelengths, the TPA of Si$_3$N$_4$ in this wavelength range is absent, which yields much higher values of nonlinear FOM for both the uncoated and GO-coated Si$_3$N$_4$ waveguides. Hence, the motivation in integrating GO films onto Si waveguides lies very much in increasing the nonlinear FOM, whereas for Si$_3$N$_4$ waveguides, the main benefit of integrating GO films is to increase the nonlinearity (i.e., nonlinear parameter $\gamma$) without introducing additional nonlinear loss.

### TABLE I.
### COMPARISON OF GO-COATED SI$_3$N$_4$ AND SI WAVEGUIDES

| Parameters | Si | Si$_3$N$_4$ |
|---|---|---|
| Refractive index [a] | 3.48 | 1.99 |
| $n_2$ [a] (m$^2$/W) | $6 \times 10^{-18}$ | $2.59 \times 10^{-19}$ |
| Waveguide dimension (μm) | $0.50 \times 0.22$ | $1.60 \times 0.66$ |
| Waveguide length (mm) | 3.0 | 20.0 |
| GO film length (mm) | 2.2 | 1.4 |
| Waveguide propagation loss (dB/cm) | 4.3 | 0.5 |
| Excess propagation loss of 1 layer of GO (dB/cm) | 20.5 | 3.0 |
| $\gamma_{WG}$ [b] (W$^{-1}$m$^{-1}$) | 288.0 | 1.5 |
| $\gamma_{hybrid}$ [c] (W$^{-1}$m$^{-1}$) | 668.0 ($N = 1$) | 11.5 ($N = 1$) |
|  | 990.0 ($N = 2$) | 27.6 ($N = 2$) |
| Fit $n_2$ of GO ($\times 10^{-14}$ m$^2$/W) | 1.42 ($N = 1$) | 1.23 ($N = 1$) |
|  | 1.33 ($N = 2$) | 1.19 ($N = 2$) |
| FOM [d] | 1.1 ($N = 1$) | >>1 |
|  | 2.4 ($N = 2$) |  |
| Ref. | [41] | This work |

[a] These values are at 1550 nm.
[b] $\gamma_{WG}$: nonlinear parameters of the bare waveguides.
[c] $\gamma_{hybrid}$: nonlinear parameters of the hybrid waveguides with 1 and 2 layers of GO.
[d] The definition of $FOM = n_2 / (\lambda\beta_{TPA})$ is the same as those in Refs. [25,27], with $n_2$ and $\beta_{TPA}$ denoting the effective Kerr coefficient and TPA coefficient of the waveguides, respectively, and $\lambda$ the light wavelength at 1550 nm.

Finally, our observation of SA in the GO-coated Si$_3$N$_4$ waveguides effectively equates to having a negative nonlinear FOM, and so the very concept and utility of introducing a FOM, as first proposed [63], arguably does not apply here. For both the Si and Si$_3$N$_4$ platforms, reducing the GO film loss further through improved fabrication and integration methods (e.g., by using GO solutions with improved purity and optimized flake sizes) will directly benefit the nonlinear performance of all devices that incorporate GO films.

## VI. CONCLUSION

We experimentally demonstrate enhanced SPM in Si$_3$N$_4$ waveguides integrated with 2D GO films. The integration of GO films is achieved by using a solution-based, transfer-free coating method with precise control of the film thickness. SPM measurements are performed using both picosecond and femtosecond optical pulses. The GO-coated Si$_3$N$_4$ waveguides show more significant spectral broadening than the uncoated waveguide, with a maximum BF of ~3.4 being achieved for a device with 2 layers of GO. The experimental results show good agreement with theory, achieving up to ~18.4 times improvement in the waveguide nonlinear parameter compared to uncoated waveguide and a fit $n_2$ of GO that is about 5 orders of magnitude higher than $Si_3N_4$. Analysis for the influence of GO film's length, coating position, and SA on the SPM performance is also provided. This work demonstrates that the $Si_3N_4$ can be effectively transformed into a highly performing CMOS-compatible nonlinear photonic platform by integrating 2D GO films.